\documentclass[twocolumn]{aastex6}

\usepackage{graphicx}
\usepackage{amssymb}
\usepackage{amsmath}
\usepackage{layout}
\usepackage{url}
\usepackage{subfigure}
\usepackage{color}
\usepackage{enumerate}
\def\beq{\begin{equation}}
\def\eeq{\end{equation}}
\def\bey{\begin{eqnarray}}
\def\eey{\end{eqnarray}}

\begin{document}

\title{Probing the Emission Mechanism and Magnetic Field of Neutrino Blazars with Multi-Wavelength Polarization Signatures}


\author{Haocheng Zhang\altaffilmark{1}, Ke Fang\altaffilmark{2,3}, Hui Li\altaffilmark{4}, Dimitrios Giannios\altaffilmark{1,5,6}, Markus B\"ottcher\altaffilmark{7}, Sara Buson\altaffilmark{8,9,10}}

\altaffiltext{1}{Department of Physics and Astronomy, Purdue University, West Lafayette, IN 47907, USA}

\altaffiltext{2}{JSI Fellow, Department of Astronomy, University of Maryland, College Park, MD 20742, USA}
\altaffiltext{3}{Einstein Fellow, Kavli Institute for Particle Astrophysics and Cosmology (KIPAC), Stanford University, Stanford, CA, 94305, USA}

\altaffiltext{4}{Theoretical Division, Los Alamos National Laboratory, Los Alamos, NM 87545, USA}

\altaffiltext{5}{Department of Physics, University of Crete, Voutes, GR-70013, Heraklion, Greece}

\altaffiltext{6}{Institute of Astrophysics, Foundation for Research and Technology Hellas, Voutes, 70013, Heraklion, Greece}

\altaffiltext{7}{Centre for Space Research, North-West University, Potchefstroom, 2531, South Africa}

\altaffiltext{8}{NASA Postdoctoral Program Fellow, Universities Space Research Association, USA\\
based at Goddard Space Flight Center, Greenbelt, MD 20771, USA}
\altaffiltext{9}{now at Institut f\"{u}r Th. Physik und Astrophysik, University of W\"{u}rzburg,  Emil-Fischer-Str. 31 D-97074, W\"{u}rzburg, Germany}
\altaffiltext{10}{University of Maryland Baltimore County, 1000 Hilltop Circle, Baltimore, MD 21250, USA}

\begin{abstract}

The characteristic two-component blazar spectral energy distribution (SED) can be of either leptonic and/or hadronic origins. The potential association of the high-energy neutrino event IceCube-170922A with the flaring blazar TXS~0506+056 indicates that hadronic processes may operate in a blazar jet. Despite multi-wavelength follow-ups of the event and extensive theoretical modelings, the radiation mechanisms and the underlying magnetic field strength and configuration remain poorly understood. In this paper, we consider generic leptonic and hadronic blazar spectral models with distinct magnetic field strengths and radiation mechanisms. We analytically reproduce the SEDs and the neutrino flux of hadronic models, and predict their X-ray to $\gamma$-ray polarization degrees. Furthermore, by performing relativistic magnetohydrodynamic (RMHD) simulations taking into account the polarization-dependent radiation transfer, we study the time-dependent multi-wavelength polarization variability of the proton synchrotron model under a shock scenario. Our results suggest that the high-energy polarization degree and the neutrino flux can be jointly used to pinpoint the leptonic and/or hadronic blazar radiation mechanisms in the X-ray and $\gamma$-ray bands, and to infer the magnetic field strength in the emission region. Additionally, the temporal multi-wavelength polarization signatures in the proton synchrotron model shed light on the jet energy composition and the dynamical importance of magnetic fields in the blazar emission region. Future multi-wavelength polarimetry facilities such as {\it IXPE} and {\it AMEGO} together with neutrino telescopes such as {\it IceCube} can  provide unprecedented observational constraints to probe the blazar radiation mechanisms and jet dynamics.

\end{abstract}

\keywords{galaxies: active --- galaxies: jets --- radiation mechanisms: non-thermal --- relativistic processes}

\section{Introduction}

Blazars are active galactic nuclei (AGN) whose jet is pointing very close to our line of sight (LOS). They are among the most powerful astrophysical objects in the universe, and exhibit highly variable emission from radio up to TeV $\gamma$-rays \citep{Ackermann16}, indicating extreme particle acceleration. As a result, blazars have long been suspected as the extragalactic sources of ultra-high-energy cosmic rays (UHECRs). Blazar SEDs are typically composed of two parts: a low-energy component from radio up to soft X-ray and a high-energy component from X-ray up to TeV $\gamma$-ray. The low-energy component is generally believed to be the synchrotron emission of primary electrons, as it usually shows a high level of polarization \citep{Scarpa97}. The origin of the high-energy component however remains debated. In a leptonic scenario, the high-energy emission is produced by the same primary electrons upscattering seed photons through the inverse Compton (IC) process. The seed photons can be the low-energy synchrotron emission by the same population of electrons \citep[synchrotron-self Compton, SSC, e.g.,][]{Marscher85,Maraschi92} or external photon fields such as the thermal radiation from the accretion disk, the broad line region (BLR), and the dusty torus \citep[external Compton, EC, e.g.,][]{Dermer92,Sikora94}. Alternatively, in a hadronic scenario, the high-energy component is produced by the primary proton synchrotron (PS) and/or the synchrotron of secondary charged particles from hadronic interactions \citep{Mannheim92a,Mucke01,Petropoulou15}. Usually the hadronic model infers the acceleration of UHECRs in the blazar zone. So far the two models result in indistinguishable SED fittings \citep{Boettcher13,Cerruti15}. Time-dependent studies of the leptonic and hadronic models have shown some moderate differences in the multi-wavelength light curves \citep{Li00,Chen14,Diltz15,Zhang16b}. However, these features are hard to be quantified in practice even with simultaneous multi-wavelength coverage. Therefore, we need additional constraints to understand the blazar emission.

The production of high-energy neutrinos is the smoking gun of cosmic hadronic interactions. If blazars accelerate a considerable amount of high-energy protons, they may interact with the local strong blazar radiation field and produce charged pions, which then decay and emit neutrinos. The recent {\it IceCube}-170922A event was reported to be coincident with the blazar TXS~0506+056 flaring state \citep{IceCube18a,IceCube18b}. Following the neutrino alert, the blazar was detected in multi-wavelength campaigns including very-high-energy $\gamma$-rays \citep{IceCube18a, MAGIC18, 2018ApJ...861L..20A}. All wavelengths exhibit strong variability, with a $\sim 7\%$ optical polarization degree during flares reported by the Kanata Telescope \citep{IceCube18a}. Theoretically, many models have been proposed to explain the SED of the TXS~0506+056 flare and the IceCube-170922A event \citep[e.g.,][]{Cerruti19,Gao19,Keivani18,MAGIC18,Reimer18}. These models can be categorized into two generic groups: a leptonic setup with a subdominant hadronic component, where the IC dominates the high-energy emission but the subdominant hadronic component produces the neutrinos and makes considerable X-rays through hadronic cascading synchrotron; and a hadronic setup where the X-ray consists of both PS and cascading synchrotron but the $\gamma$-ray is dominated by the PS. Nonetheless, even though the two scenarios imply very different physical conditions (such as magnetic field strength) in the emission region, the above theoretical studies cannot unambiguously pinpoint the high-energy radiation mechanisms. In addition, these studies generally find that a simple one-zone model cannot fully explain the temporal behavior of the radiation and neutrino emission. Thus the underlying particle acceleration and the related magnetic field evolution are poorly constrained.

Magnetic field strength and evolution hold the key to disentangle various neutrino blazar models and shed light on the particle evolution. Polarimetry has been a standard probe of the blazar magnetic field evolution. In particular, the optical polarization signatures have been found highly variable alongside multi-wavelength flares, including significant polarization degree drops and $\gtrsim 180^{\circ}$ polarization angle swings \citep[e.g.,][]{Abdo10,Angelakis16}. Numerical simulations have shown that the amplitude of the optical polarization variability may indicate the magnetic energy percentage (often called the magnetization) in the blazar zone \citep{Zhang16a,Zhang17}. In the high-energy bands, the PS scenario predicts a high polarization degree in X-rays and $\gamma$-rays \citep{Zhang13,Paliya18}, which may be detected by future high-energy polarimeters such as IXPE and AMEGO \citep{Weisskopf16,McEnery17}. The multi-wavelength polarimetry along with neutrino detection may unveil the details of the jet dynamics, particle acceleration, and radiation mechanisms.

In this paper, we make the first attempt to predict the multi-wavelength polarization signatures of both leptonic and hadronic blazar models, via semi-analytical calculation of broadband spectral polarization degree and first-principle-integrated polarized radiation transfer simulation of temporal polarization signatures. We aim to identify generic observable features that can be directly confronted with future multi-messenger and multi-wavelength polarimetry observations of blazars. In section 2, we take the TXS~0506+056 SED as an example, and apply one-zone spectral modeling of four benchmark leptonic and leptohadronic scenarios that correspond to drastically different jet physical conditions. Based on the fitting parameters, we predict the X-ray to MeV $\gamma$-ray spectral polarization degree that can be directly confronted with future high-energy polarimeters. In section 3, we perform 3D RMHD simulations with polarized radiation transfer to study the effects of the blazar zone magnetization on the temporal behavior of multi-wavelength polarization signatures of a PS blazar model. We demonstrate that the optical and MeV $\gamma$-ray polarization patterns can be very similar, and drastic temporal polarization variations, such as a polarization angle swing, indicate strong shock compression in a weakly magnetized emission region. We discuss the implications of our results on current and future observations in section 4, and summarize in section 5.

\section{High-Energy Spectral Polarization Degree}

In this section, we compute the high-energy spectral polarization degrees of four generic leptonic and leptohadronic blazar spectral models. Previous works have shown that the magnetic field is a key parameter that affects the contribution of different radiation mechanisms to the blazar high-energy spectral component \citep[e.g.,][]{Boettcher13,Cerruti19,Reimer18}. We thus pick four benchmark parameter regimes with different radiation mechanisms and physical properties, especially the magnetic field strength and magnetization.
To facilitate realistic predictions of the high-energy polarization, we take the recent TXS~0506+056 observation \citep{IceCube18a} to derive the general physical parameters and particle distributions via one-zone spectral modeling. While we do not aim at detailed fitting, the derived parameters can help to illustrate their effects on the high-energy SED and spectral polarization. In the following, we first describe our models, then compute the broadband SEDs and neutrino flux, finally evaluate the spectral polarization degree.

\subsection{Model Description}

\begin{table*}[t]
\scriptsize
\parbox{\linewidth}{
\centering
\begin{tabular}{|l|c|c|c|c|}\hline
\multicolumn{5}{|c|}{Model Parameters}                                          \\ \hline
Fitting Parameters                   & Model 1 & Model 2 & Model 3 & Model 4    \\ \hline
Redshift ($z$)                       & \multicolumn{4}{c|}{0.3365}              \\ \hline
Bulk Lorentz factor ($\Gamma_{obs}$) & \multicolumn{4}{c|}{10}                  \\ \hline
Viewing angle ($\theta_{obs}$)       & \multicolumn{4}{c|}{0}                   \\ \hline
Blob size ($R'$, cm)                 & \multicolumn{4}{c|}{$5\times 10^{15}$}   \\ \hline
Escape time ($t'_{esc}$, s)          & \multicolumn{4}{c|}{$6.67\times 10^{5}$} \\ \hline
Magnetic field ($B'$, G)             & 1.3     & 1.2     & 50    & 100          \\ \hline
Electron minimal Lorentz factor ($\gamma'_{e,1}$) & 2000 & 4000 & 400 & 250     \\ \hline
Electron maximal Lorentz factor ($\gamma'_{e,2}$) & 40000 & 40000 & 20000 & 10000 \\ \hline
Electron power-law index ($p_e$)     & 2.0 & 2.0 & 3.2 & 3.2                    \\ \hline
Electron kinetic luminosity ($L_{e,kin}$, $10^{43}~\rm{erg\,s^{-1}}$) & 40 & 32 & 6 & 1.5 \\ \hline
Proton minimal Lorentz factor ($\gamma'_{p,1}$)  & -- & \multicolumn{3}{c|}{1}  \\ \hline
Proton turnover Lorentz factor ($\gamma'_{p,b}$) & -- & $2\times 10^8$ & $2\times 10^8$ & -- \\ \hline
Proton maximal Lorentz factor ($\gamma'_{p,2}$)  & -- & \multicolumn{3}{c|}{$2\times 10^9$} \\ \hline
Proton power-law index before turnover ($p_{p,1}$) & -- & 2.1 & 2.1 & 2.2       \\ \hline
Proton power-law index after turnover ($p_{p,2}$)  & -- & 3.0 & 3.0 & --        \\ \hline
Proton kinetic luminosity ($L_{p,kin}$, $10^{46}~\rm{erg\,s^{-1}}$) & -- & 3.2 & 8 & 9 \\ \hline
\multicolumn{5}{|c|}{Derived Quantities}                                        \\ \hline
Jet Power ($P^{obs}_{jet}=L_{e,kin}+L_{p,kin}+L_B$, $10^{46}~\rm{erg\,s^{-1}}$) & 0.0072 & 3.2 & 10.3 & 18.4 \\ \hline
$\sigma'$                            & 0.28    & 0.0004  & 0.29  & 1.04        \\ \hline
$L_{e,kin}/L_{p,kin}$                & --      & $1.2\times10^{-3}$ & $8.1\times10^{-6}$ & $2.7\times10^{-6}$ \\ \hline
\end{tabular}}
\caption{Model parameters and derived quantities of the four spectral fitting models shown in Figure \ref{figure:sed}. The redshift measurement of TXS~0506+056 is taken from ~\citet{Ajello14}. \label{table:sed}}
\end{table*}
The key difference of the IC-dominated and PS-dominated blazar models generally lies in the magnetic field strength. In the former case, the magnetic field is usually on the order of $0.1~\rm{G}$, so that for most blazars the seed photon energy density for IC is comparable to or even larger than the magnetic energy density. Additionally, even if there exists an energetic proton population, they can only cool through pion production processes. Therefore, the high-energy emission is mostly dominated by the IC. In contrast, PS-dominated models usually require a magnetic field of at least $10~\rm{G}$. The IC seed photon energy density is then negligible compared to the magnetic energy density in most blazars, and protons can emit synchrotron efficiently. Given the above considerations, we set up the following four scenarios.
\begin{enumerate}
    \item A pure IC-dominated model, where the high-energy spectral component is dominated by electron SSC. In principle, there could be a negligible hadronic component.
    \item An IC-dominated model, where the high-energy spectral component is dominated by SSC, but it contains a small contribution from the hadronic component.
    \item A PS-dominated model, where the high-energy spectral component is dominated by synchrotron of both protons and hadronic cascading pairs. There could be a small contribution from SSC of electrons as well.
    \item A pure PS-dominated model, where the high-energy spectral component is dominated by proton synchrotron, with a minor contribution from the cascading pair synchrotron. Primary electron SSC is negligible.
\end{enumerate}
Since TXS~0506+056 does not present a strong thermal component in the SED, we will not include a thermal background in our spectral fitting. However, in principle a small thermal flux could hide below the primary electron synchrotron, and may contribute to the photopion processes and the external Compton emission (EC).

One can immediately derive some general conclusions from the above four models. Model 1 is unlikely to produce any detectable neutrinos. Several previous works have found that Model 2 provides the best fit to the electromagnetic and neutrino emission from TXS~0506+056, although the modelled neutrino flux is significantly lower than the uppler limit inferred by {\it IceCube-170922A} \citep[e.g.,][]{Cerruti19}. 
The major difference between Model 3 and 4 lies in the magnetic field strength, which results in different cooling rates. 

In the following all quantities in the comoving frame of the blazar jet are marked with a prime, those in a frame that is stationary to the black hole are  unscripted,  and those in the observer's frame are marked with a subscript ``obs". 

Assuming that the blazar jet is relativistic with a bulk Lorentz factor $\Gamma_{obs}= 10$, and the viewing angle is at $\theta_{\rm obs}=1/\Gamma_{obs}$, one can find that the observed neutrino energy of $E_{\nu}\sim 300~\rm{TeV}$ requires a parent proton energy of 
\beq
\gamma_p' \sim 10^6\,\delta_1^{-1}~~, 
\eeq
where $\delta_{obs}= [\Gamma_{\rm obs}(1-\beta_{\rm obs}\cos\theta_{\rm obs})]^{-1} = 10\delta_1$ is the Dopper factor, with $\beta_{\rm obs}$ being the speed of the jet in unit of $c$. Then the target photon energy that corresponds to the $\Delta$-resonance of the photopion production is 
\beq
\epsilon'_{p\gamma} \approx \frac{\bar{\epsilon}_\Delta}{2\gamma_p'} = 150\,\gamma_{p,6}'^{-1}~~\rm eV,
\eeq
where we take $\gamma'_p=10^6 \gamma'_{p,6}$ and $\bar{\epsilon}_\Delta\sim 0.3$~GeV is the resonance energy of the photopion production. The cooling rates of the photopion and synchrotron process are given by
\begin{equation}
\begin{array}{l}
\dot{\gamma}'_{p,p\gamma}=-c  \sigma_{p\gamma}  \frac{u'_{rad}}{m_ec^2}\gamma_p'\\
\dot{\gamma}'_{p,syn}=-\frac{4}{3}c\sigma_T\frac{u'_B}{m_e c^2}(\frac{m_e}{m_p})^3 \gamma_p^{\prime 2}
\end{array}~~,
\end{equation}
respectively, where $\sigma_{p\gamma}\sim 70\,\mu$b is the inelasticity-weighted average photopion cross section, $\sigma_T$ is the Thomson cross section, $u'_{\rm rad}$ and $u'_B$ are the photopion target photon energy density and magnetic energy density, respectively. Then one can find that
\begin{equation}
\frac{\dot{\gamma}'_{p,p\gamma}}{\dot{\gamma}'_{p,syn}}\sim 7\times 10^5~\frac{u'_{rad}}{u'_B}\gamma_p^{\prime -1}~~.
\end{equation}
The target photon energy density at $\epsilon'_{p\gamma}$ is not well constrained by observations, but previous models have shown that an external radiation field is necessary in order to produce an observable  amount of neutrinos \citep{Cerruti19,Reimer18}. If we adopt  $u'_{rad}\sim 100~\rm{erg\,cm^{-3}}$ as suggested by \cite{Reimer18}, then we can easily find that for moderate magnetic field strength, $B'\lesssim 50~\rm{G}$, the photopion cooling rate is faster than the synchrotron for $\gamma'_p\sim 10^6$, leading to efficient neutrino production. 

\subsection{Broadband SED}
\begin{figure*}
\centering
\includegraphics[width=\linewidth]{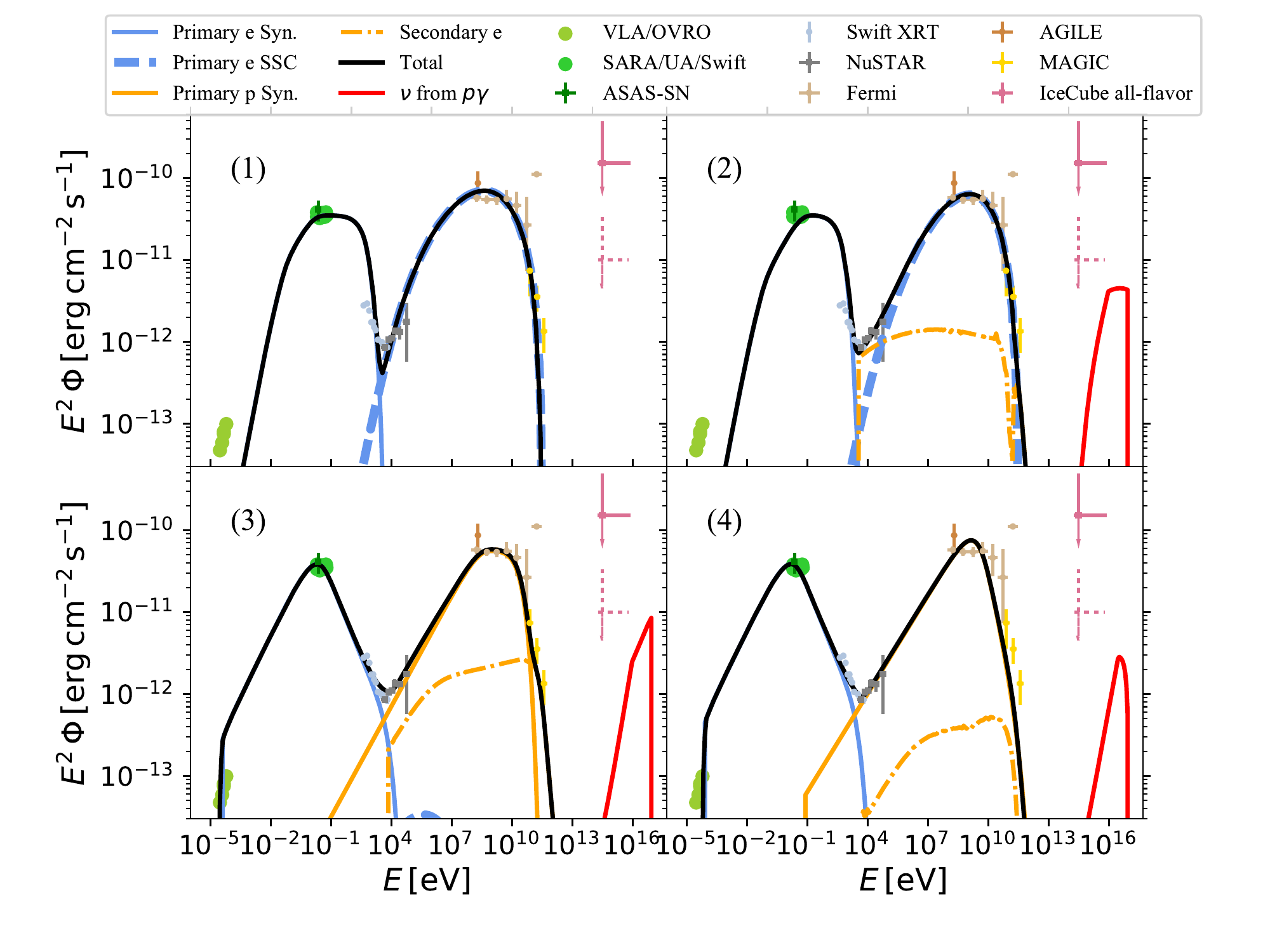}
\caption{Multi-wavelength spectral fitting using the inverse Compton scenario (model 1 and 2, upper panels) and the proton synchrotron scenario (model 3 and 4, lower panels) comparing to the TXS~0506+056 observation \citep{IceCube18a}. The model parameters are listed in Table~\ref{table:sed}. In all panels, the black solid curves are the total SED, the blue curves represent the synchrotron emission (solid) and inverse Compton emission (thick dashed) by primary electrons, orange curves correspond to photons emitted by primary protons (solid) and their secondaries (dash-dotted), and red solid curves show neutrinos from the photopion production of primary protons. \label{figure:sed} 
}
\end{figure*}
Our spectral model is based on  a stationary lepto-hadronic radiation code developed by \cite{Boettcher13}, which semi-analytically treats radiative, photomeson, and adiabatic cooling as well as particle escaping. To facilitate direct comparisons, we choose the same bulk Lorentz factor, viewing angle, and emission region size for all four models. The default particle escaping time scale is chosen as 4 times of the light crossing time scale. The derived SED is corrected for the extragalactic background light attenuation. Figure \ref{figure:sed} shows our fitting results with parameters listed in Table \ref{table:sed}. We remind the readers that we do not aim at accurate fitting of the TXS~0506+056 observation. Instead, we are interested in placing  general constraints on the parameter space of the four models, so that their key differences can stand out.

From Figure \ref{figure:sed}, we can identify the dominating radiation mechanisms in the high-energy bands for the four models. In Model 1, the primary electron SSC emission dominates the entire high-energy spectral component. Model 2 is similar to Model 1 but it has a significant synchrotron contribution by pairs in the X-ray band. These pairs are the byproducts of the photopion process, where the energetic protons interact with the primary electron synchrotron and produce neutrinos. We remind the readers that in the broadband SED calculation we do not include any external photon field for Compton scattering or hadronic interactions, because TXS~0506+056 is often considered as a BL Lac \citep[however, some other works suggest that it could be an FSRQ, most recent one refer to][]{Padovani19}. Generally speaking, however, the EC contribution can be significant in the $\gamma$-ray bands, especially for FSRQs. Both Model 3 and 4 have primary proton synchrotron dominating the high-energy bands, but in Model 3 the magnetic field is weaker. Since we do not include an external photon field, the target photon field for photomeson processes in Model 2-4 is the primary electron synchrotron emission.

\subsection{Neutrino Flux \label{sec:fpgamma}}

High-energy neutrinos are produced when primary protons interact with ambient photons in the jet. Using the measurement of the optical to X-ray flux of TXS~0506+056 we can derive the energy density of the target photons and the effective optical depth for the $p\gamma$ interaction (also referred as the pion production efficiency, $f'_{p\gamma}$ below).

If neutrinos and $\gamma$ rays are produced in the same region, and target photons come from an optically thin area such as the emission region itself, we can obtain its density from the broadband SED. The pion production efficiency can be approximated as   
 
\bey\label{eqn:f_pg_obs}
f'_{p\gamma} &\sim&\left[ \frac{\left(\epsilon\,F_\epsilon\right)^{\rm obs}\, 4\pi d_L^2 }{\delta^4\,4\pi R'^2 c \,\epsilon'}\right] \sigma_{p\gamma} c\,t'_{\rm esc} \\ \nonumber 
&=&1.8\times10^{-5}\,\delta_1^{-4}\,\left[\frac{\left(\epsilon F_\epsilon\right)^{\rm obs}}{10^{-12}\,\rm erg\,cm^{-2}\,s^{-1}}\right]\left(\frac{\epsilon'}{200\,\rm eV}\right)^{-1}, 
\eey
The fiducial value $\epsilon'=200$~eV is chosen  to be the target photon energy that yields the neutrino emission in the energy range observed by IceCube from TXS~0506+056.  Assuming this target photon energies, the photopion production via the $\Delta$-resonance implies an emission of neutrinos with energies of $\sim 290$~TeV in the observer frame. $R'$ and $t'_{\rm esc}$ are the size of the emission region and the escape time of protons respectively. Their values are presented in Table~\ref{table:sed}. 

Neutrinos carry $\sim 3/4$ of the energy of the charged pions, and about $1/2$ of the $p\gamma$ interactions lead to the production of charged pions. The neutrino flux can thus be estimated by 
\beq\label{eqn:neuFlux}
\left(\epsilon_\nu^2F_{\epsilon_\nu}\right)^{\rm obs} \approx \frac{\delta^4}{4\pi d_L^2} \frac{3}{8} f_{p\gamma}' \gamma_p'^2 m_p c^2 \frac{d \dot{N'}}{d\gamma_p'}\big|_{E_p' \approx 20\,\epsilon_\nu (1+z) / \delta}
\eeq
where $d\dot{N}'/d\gamma'_p$ is the proton injection rate derived from Tabel~\ref{table:sed}. 
The neutrino flux of the four models is shown in Figure~\ref{figure:sed}. Apparently, if the neutrinos are produced within the multi-wavelength flaring region, their flux should be much lower than the observed level, consistent with similar calculations done for example by \citealt{Keivani18}.
 
To explain both the neutrinos and electromagnetic flaring emission observed at all wavelengths, below we also consider a two-zone model. Previous works have suggested that TXS~0506+056 may not be a typical BL Lac \citep[e.g.,][]{Padovani19}, but it may have a weak external thermal component from the broad line region (BLR). Here the two zones refers to within the BLR (including the thermal emission from BLR, zone 1) and beyond BLR (without thermal emission from BLR, zone 2). Notice that this is the only place in the paper that we consider the contribution of external photon field, so as to explain neutrino production. We envision that the relativistic jet, containing high-energy protons and electrons, passes through the BLR and continues to move away from the central engine. In zone 1, the highly energetic protons interact with the dense BLR photon field and produce neutrinos through photomeson processes. The $\sim 100~\rm{GeV}$ $\gamma$-rays, however, can hardly escape during the neutrino production phase due to the large optical depth. Thus we do not expect to observe the  $\sim 100~\rm{GeV}$ $\gamma$-ray emission accompayning the neutrino production. When these electrons and protons move to zone 2, the BLR photon field becomes negligible, and $\sim 100~\rm{GeV}$ photons start to escape. However, in zone 2 neutrino production is greatly reduced due to the lack of target photon field. Therefore, this transition through the two zones predicts a delay of the $100~\rm{GeV}$ $\gamma$-ray emission with respect to the neutrino event, while the radiation emitted at energies lower than $\sim 10~\rm{GeV}$ should appear simultaneous with the neutrino event. 

Such two-zone model is appealing, even though it introduces several free parameters and diminish the model prediction power. Here we choose to estimate the neutrino production rate of this model by using the same physical parameters for both zones as in the above one-zone model, and introduce a BLR photon field in zone 1. The energy density of the line photons in BLR is $u_{\rm BLR} \approx  0.3 \,\rm erg\,cm^{-3}$ in the black hole frame \citep{GT08}. The most prominent contribution comes from the Ly$\alpha$ at $\epsilon_{\rm UV}=2\times 10^{15}$~Hz. The photopion interaction rate in the comoving frame is ${t'}^{-1}_{p\gamma} = \sigma_{p\gamma}\,c\,\Gamma^2\,u_{\rm BLR}/(\Gamma\,\epsilon_{\rm UV})$. The dynamical time for the jet to travel through the external field is $t'_{\rm dyn} = R / \Gamma\,c$, thus the interaction efficiency is
\beq
f'_{p\gamma} = \frac{t'_{\rm dyn}}{t'_{p\gamma}} = 1.4\times10^{-2}\,R_{\rm BLR, 16} 
\eeq
where $R_{\rm BLR}$ is the size of the BLR. This shows that the introduction of a BLR photon field indeed allows a higher photopion production efficiency that may explain the IceCube event.

To summarize, if neutrinos and VHE $\gamma$-rays are produced co-spatially, we expect that the actual neutrino flux level should be much lower than the estimated value suggested by \cite{IceCube18a, IceCube18b}. But if they are produced in different regions, a strong external photon background would be allowed. The pion production efficiency could be greatly enhanced, leading to an average neutrino flux that is comparable to the IceCube measurement. However, in this model, the VHE $\gamma$-ray flare should appear later than the neutrino emission. 

\subsection{X-ray to $\gamma$-ray Polarization Degree}

Given the fitting parameters and the particle distributions, we calculate the spectral polarization degree in the high-energy spectral component. We follow the procedure developed by \cite{Zhang13} to evaluate the spectral polarization. The calculation assumes a mono-directional magnetic field that is perpendicular to the LOS in the comoving frame of the emission region, and calculate the polarization-dependent radiation power parallel and perpendicular to the magnetic field for primary electron SSC, as well as synchrotron by primary electrons, primary protons, and secondary pairs from hadronic interactions. The evaluation is done by integration of the derived particle spectra from the SED fitting, thus it naturally includes all spectral changes and breaks derived from the fitting. Generally speaking, since the electrons are distributed isotropically in the comoving frame, the EC is unpolarized. In the end, by taking the EC component as unpolarized, the procedure obtains the maximal spectral polarization degree in all high-energy bands.

In practice, the magnetic field in the emission region is partially ordered, thus we need to correct the spectral polarization degree under a partially ordered magnetic field. \cite{Zhang16b} have simulated the multi-wavelength polarization signatures in a helical magnetic field, concluding that the partially ordered magnetic field should lower the polarization degree in the optical and high-energy bands by the same ratio. \cite{Paliya18} have used a simple approximation to include this effect, and we will follow it here. The general formalism for the observed polarization degree is
\begin{equation}
\Pi(\nu')=Z_m\frac{\Pi_{max}(\nu')P'_{pol}(\nu')}{P'_{pol}(\nu')+P'_{un}(\nu')} ~~,
\end{equation}
where $Z_m$ is the correction factor for the partially ordered magnetic field, $\Pi_{max}$ is the maximal polarization degree, $P'_{pol}$ and $P'_{un}$ are the radiation powers of polarized components (such as SSC and synchrotron) and unpolarized components (such as EC and thermal), respectively, in a mono-directional magnetic field. Since we ignore all thermal and EC components in our spectral modeling, then $\Pi(\nu')=Z_m\Pi_{max}(\nu')$ for both optical and high-energy bands. Assuming a default observed optical polarization degree of $10\%$, then we find
\begin{equation}
\Pi(\nu')=10\% \frac{\Pi_{max}(\nu')}{\Pi_{max,o}}~~,
\end{equation}
where $\Pi_{max,o}$ is the maximal optical polarization degree, which can be calculated based on the fitting parameters.

\begin{figure}[hbt]
    \centering
    \includegraphics[width=0.5\textwidth]{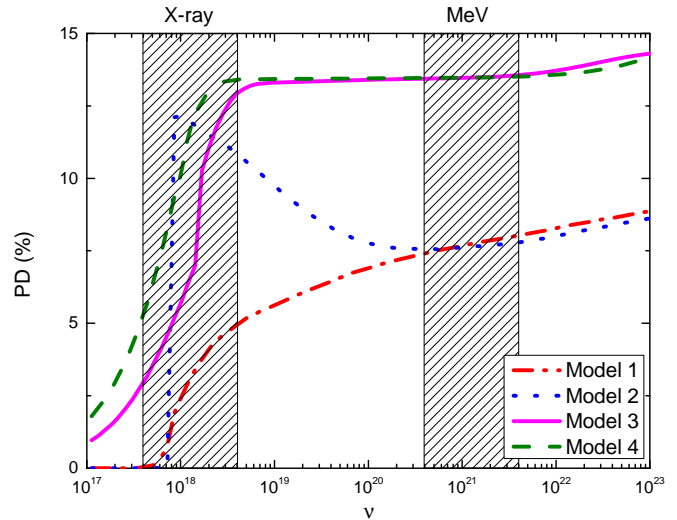}
    \caption{The X-ray to $\gamma$-ray spectral polarization degree based on the spectral fitting of TXS~0506+056 observation. The shaded regions correspond to the X-ray and MeV $\gamma$-ray bands. The spectral polarization degree is calculated based on the fitting results in Figure \ref{figure:sed} with parameters listed in Table \ref{table:sed}. \label{figure:specpol}}
\end{figure}

Figure \ref{figure:specpol} shows the predicted spectral polarization degree from X-ray to $\gamma$-ray energies for all four models. Since the SSC generally lowers the seed synchrotron polarization degree by half, Model 1 predicts a low polarization degree, $\sim 7\%$, throughout the high-energy spectral component. The sharp rise of the polarization degree in Model 2 is due to the strong secondary pair synchrotron in the X-ray bands. The polarization degree gradually drops towards higher energies, as the dominating radiation process transitions into the SSC. We want to emphasize that if the blazar has a strong EC component in the $\gamma$-ray bands, we expect that Model 1 and 2 should appear more weakly polarized or even unpolarized towards higher energies, as the EC component is unpolarized. Although Model 3 has a strong secondary pair synchrotron in the X-ray to soft $\gamma$-ray bands, 3 and 4 predict very similar high-energy polarization degree at $\sim 15\%$ level throughout the high-energy spectral component, except for the rising part in the X-ray bands. This rising part marks the transition where the proton emission processes become dominating in the high-energy spectral component, which is dependent on the exact fitting parameters for the PS component. The $\sim 15\%$ polarization degree for both models is easily understandable, as typically all synchrotron emission is similarly polarized, regardless of the type of the charged particles. We notice that generally all spectral polarization curves tend to rise towards higher energies. This is because the underlying particle spectra become softer at higher energies due to cooling effects, which boost the polarization degree. Clearly, PS-dominated models generally show higher polarization degree than IC-dominated models.

\section{Temporal Multi-Wavelength Polarization Signatures}

In this section, we study the effects of magnetization of the blazar emission region on the temporal behavior of multi-wavelength polarization signatures. As shown in the previous section, the PS models predict considerably higher X-ray and $\gamma$-ray polarization, which may be more interesting and easier to detect with future high-energy polarimeters such as IXPE and AMEGO. We limit our simulations in the PS setups. We consider the archetypal blazar flaring model where a shock propagates through the emission region and triggers strong particle acceleration. We choose two emission region setups, namely, a low magnetization environment with the magnetization factor $\sigma'\sim 0.1$ (this factor is defined as the ratio of magnetic energy density over enthalpy in the fluid frame), and a high magnetization environment with $\sigma'\sim 1$; everything else remain the same. To evaluate the time-dependent polarization signatures, we employ 3D RMHD simulations to self-consistently evolve the magnetic field during the shock propagation. In addition, we use 3D polarized radiation transfer simulations to calculate the observed polarization signatures. In the following, we first describe our simulation setup, then present the RMHD simulation results, finally calculate the temporal polarization signatures.

\subsection{Simulation Setup}

Our general simulation setup is very similar to \cite{Zhang16a}, except that now we include the PS in the radiation processes. Here we summarize the key model setups that are directly related to our simulations. Our model assumes that the blazar jet is traveling with a bulk Lorentz factor of $\Gamma_{obs}$ in the observer's frame. In the comoving frame of the jet, the blazar emission region is initially cold (thermal pressure is trivial), pervaded by a helical magnetic field with a magnetization factor of $\sigma'$. For simplicity, we assume that the initial magnetic field is under force balance with no pre-existing turbulence, and the plasma density and thermal pressure are uniform in the emission region. A flat shock of bulk Lorentz factor $\Gamma'$ then propagates through the jet and encounters the emission region, which changes the plasma physical conditions through its interaction with the emission region and triggers local particle acceleration. After the shock leaves the emission region, we continue to simulate the jet evolution until when it reaches a quasi-stationary state. In this way, all the plasma physical quantities, in particular the evolution of the magnetic field and jet energy composition, are self-consistently simulated under first-principle magnetized fluid dynamics.

To produce the aforementioned model we use the 3D RMHD code LA-COMPASS \citep{Li03}. We perform the RMHD simulation in the jet comoving frame, with uniform Cartesian grids. The emission region is set to be a cubic box of length $5L'_0$ at the center of the simulation domain (thus, $x,~y,~z$ range from $-2.5L'_0$ to $2.5L'_0$). The shock is set to be a flat layer (thickness $0.5L'_0$) of fast-moving plasma initially with $\Gamma'=5$. To avoid any artificial effects at the injection of this fast-moving layer, we put it at $-7.5L'_0$ and let it propagate towards the emission region, thus its entering speed into the emission region can be slower. We pick the outflow boundary condition for the simulation domain. However, to get rid of any boundary effects, we choose a sufficiently large domain, $x,~y$ go from $-5L'_0$ to $5L'_0$, and $z$ goes from $-20L'_0$ to $20L'_0$, so that throughout our simulation, no signal from the emission region can arrive at the boundary. Within the emission region, we take a force-free helical magnetic field morphology in the form of 
\begin{equation}
\begin{array}{l}
B'_z = B'_0\times J_0(kr') \\
B'_{\phi} = B'_0\times J_1(kr')
\end{array}~~,
\label{equation:rmhdmag}
\end{equation}
where $J_0$ and $J_1$ are the Bessel functions of the first kind, $r'$ is the radius measured from the $z$-axis, $B'_0$ and $k$ are normalization factors of the magnetic field strength and the radius, respectively. The initial $\sigma'$ is defined as the volume average magnetization factor in the emission region, with the low magnetization setup as $\sigma'\sim 0.1$ and high magnetization setup as $\sigma'\sim 1$. Given the above magnetic field morphology, the local magnetization factor is larger than the initial $\sigma'$ at the central $z$-axis, and gradually drops outwards. In order to capture the detailed temporal and spatial evolution of the fluid physical quantities, we take a spatial resolution of $128\times 128 \times 512$ in $x,~y,~z$ directions, with an output of plasma density $\rho'$, velocity $\boldsymbol{v}'$, magnetic field $\boldsymbol{B}'$, and thermal pressure $P'$ at every grid point in every $0.5t'_0$ time step. Table \ref{table:RMHD} lists the conversion between code units and physical units.

RMHD simulations, however, do not include nonthermal particle acceleration and evolution. A detailed description of the nonthermal particle acceleration and cooling in our hadronic model is beyond the scope of this paper. For simplicity, we adopt the following assumptions to approximate the particle evolution. Given that many blazars can still emit radiation during quiescent states, we assume that the emission region has two populations of nonthermal particles for both electrons and protons. One population is a fixed and steady particle distribution $n'_{e,b}$ and $n'_{p,b}$, which is distributed uniformly in space. The underlying assumption is that this population of particles have reached equilibrium between secondary Fermi acceleration and radiative cooling. The other originates from the shock acceleration of fresh nonthermal particles, $n'_{e,s}$ and $n'_{p,s}$, which should be related to the time-dependent local shock properties. For both populations, we assume that the particle spectral distributions are of a power-law shape with an exponential cutoff,
\begin{equation}
\begin{array}{l}
n'_e(\gamma'_e)=n'_{e,0}\gamma_e^{\prime -2.0}e^{-\frac{\gamma'_e}{10^4}} \\
n'_p(\gamma'_p)=n'_{p,0}\gamma_p^{\prime -2.0}e^{-\frac{\gamma'_p}{5\times 10^8}}
\end{array}~~,
\end{equation}
where $n'_{e,0}=n'_{e,b}+n'_{e,s}$ and $n'_{p,0}=n'_{p,b}+n'_{p,s}$ are normalized by the local physical quantities. Previous simulations have shown that the light curves and polarization signatures are mostly affected by the nonthermal particle density rather than the spectral shape \citep{Zhang16a,Zhang18}, hence we fix the spectral index throughout our simulations. For $n'_{e,s}$, given the very high magnetic field in the hadronic model, the electron cooling is very fast compared to the shock propagation or light crossing time scales. Therefore, we inject $n'_{e,s}$ that is proportional to the local kinetic energy at the shock front in each RMHD time step, then in the next RMHD output time step, remove them and inject new $n'_{e,s}$ based on the refreshed local kinetic energy and the shock front location. In this way, we mimic the shock acceleration of electrons and their fast cooling. The proton injection $n'_{p,s}$, however, needs a more careful treatment. Even the high-energy protons that emit $\gamma$-rays ($\gamma'_p\sim 10^8$) have cooling time scales comparable to the light crossing time scale of the emission region. Therefore, the protons cannot be simply refreshed at each RMHD output time step. Given the very long cooling time scale, we assume that the nonthermal proton energy density $n'_{p,s}$ is proportional to the local thermal energy density in the emission region. \cite{Zhang16b} have demonstrated that if the proton cooling time scale is comparable to the light crossing time scale, as in our simulation setup, this approximation can work. However, in the case that the light crossing time scale is even shorter, the detailed proton cooling physics must be carefully evaluated, which is beyond the scope of this paper.

Finally we calculate the time-dependent polarization signatures based on the above magnetic field evolution and particle distributions. The most important physical effect here is the light crossing time scale. We use the 3DPol code to include all light crossing time effects for both electron and proton synchrotron emission \citep{Zhang16b}. The light crossing time effect is a physical effect that photons from different location in the emission region may arrive to the observer at different time, which leads to various arrival time delays. The 3DPol code takes Cartesian coordinates as in the RMHD simulation, and evaluate Stokes parameters in the comoving frame at each grid point at each time step. Then it applies ray-tracing method to propagate the beam to the plane of sky. There it adds up all the beams that arrive at the same time, and Doppler boosts all radiation and polarization signatures to the observer's frame. Due to the relativistic aberration, for blazars observed at $\theta_{obs}\sim 1/\Gamma_{obs}$ from the jet axis, in the comoving frame we are observing at $\theta'\sim 90^{\circ}$. For all simulations we take the LOS at $\theta'=90^{\circ}$, hence the Doppler factor is $\delta_{obs}=\Gamma_{obs}$. We define the polarization angle $PA=0$ when the electric vector is perpendicular the jet axis.

\begin{table}[ht]
\scriptsize
\parbox{0.49\textwidth}{
\centering
\begin{tabular}{|l|c|c|}\hline
RMHD Parameters  & Code Value          & Physical Value                       \\ \hline
Length           & $L'_0$              & $2.0\times10^{16}~\rm{cm}$           \\ \hline
Time             & $L'_0/c$            & $3.33\times10^5~\rm{s}$              \\ \hline
Velocity         & $c$                 & $3\times10^{10}~\rm{cm\,s^{-1}}$     \\ \hline
Magnetic Field   & $B'_0$              & $100~\rm{G}$                         \\ \hline
Thermal Pressure & $B_0^{\prime 2}/(4\pi)$     & $7.96~\rm{erg\,cm^{-3}}$             \\ \hline
Plasma Density   & $B_0^{\prime 2}/(4\pi c^2)$ & $8.84\times10^{-19}~\rm{g\,cm^{-3}}$ \\ \hline
\multicolumn{3}{c}{}\\
\end{tabular}}
\parbox{0.49\textwidth}{
\centering
\begin{tabular}{|l|c|}\hline
3DPol Parameters                               & Value            \\ \hline
Bulk Lorentz factor $\Gamma_{obs}$             & $10$             \\ \hline
LOS direction $\theta'$                        & $90^{\circ}$     \\ \hline
Electron minimal Lorentz factor $\gamma'_{e,1}$ & $1$              \\ \hline
Electron maximal Lorentz factor $\gamma'_{e,2}$ & $10^4$           \\ \hline
Electron power-law index $p_e$                 & $2.0$            \\ \hline
Proton minimal Lorentz factor $\gamma'_{p,1}$   & $1$              \\ \hline
Proton maximal Lorentz factor $\gamma'_{p,2}$   & $5\times 10^8$   \\ \hline
Proton power-law index $p_p$                   & $2.0$            \\ \hline
\end{tabular}}
\caption{Summary of parameters. Top: Conversion between the RMHD code units and the physical value. Bottom: Additional parameters used in the 3DPol simulation. All parameters in this table are in the comoving frame of the emission region, except for the bulk Lorentz factor. \label{table:RMHD}}
\end{table}

\subsection{Shock Interaction with Magnetic Field}

\begin{figure*}[ht]
\parbox{\linewidth}{
\centering
\includegraphics[width=\linewidth]{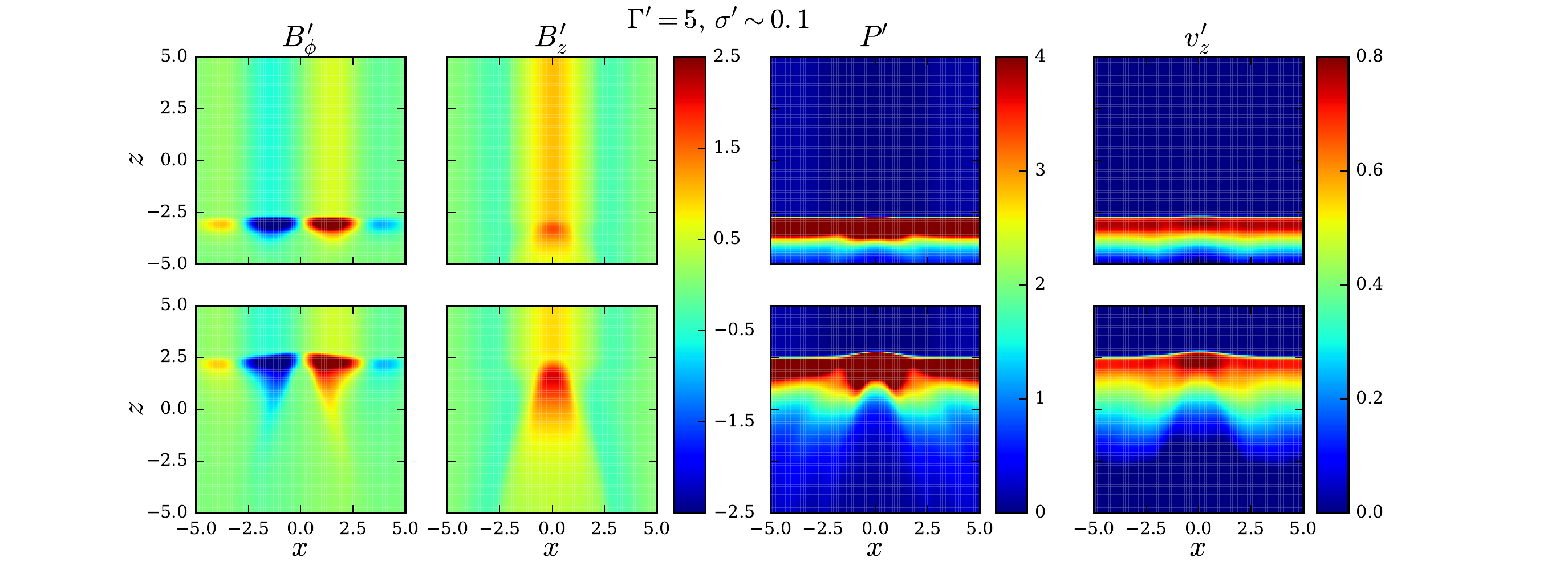}
\includegraphics[width=\linewidth]{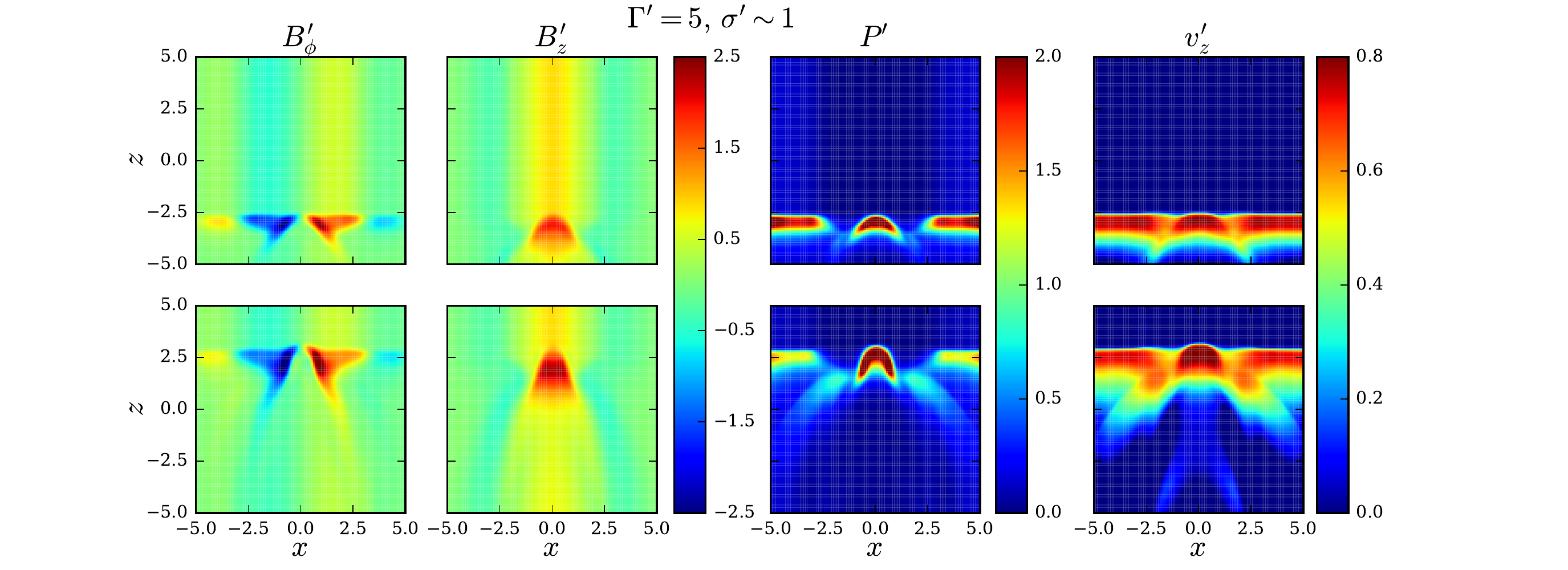}
}
\caption{RMHD simulations of the shock propagation through the emission region. Upper two rows are for the low magnetization environment, while the lower two rows are for the high magnetization environment. The left two columns are the toroidal and poloidal magnetic field strength, respectively. The third column is the thermal pressure, and the right column is the velocity in $z$ direction. For both low and high magnetization plots, the upper panels are at the time when the shock just enters the emission region, while the lower panels are when the shock is about to leave the emission region. All quantities are shown in the comoving frame with code units as shown in Table \ref{table:RMHD}. Notice that in the low and high magnetization plots, all physical quantities share the same colorbars, except for the thermal pressure, where the high magnetization colorbar has a smaller maximal value. \label{figure:RMHD}}
\end{figure*}

When the shock passes through the emission region, we observe that shocked plasma shows higher pressure and density due to compression (Figure \ref{figure:RMHD} third column). In ideal RMHD, since the magnetic field lines are frozen in the plasma, the toroidal magnetic field component is also enhanced in the shocked plasma. Given the helical magnetic field setup in Equation \ref{equation:rmhdmag}, the toroidal component is at its maximum at radius $r'\sim 1.2$ from the $z$-axis, thus we observe strengthened toroidal component near this radius (Figure \ref{figure:RMHD} first column). An interesting feature is that the poloidal component is stronger at the central jet (Figure \ref{figure:RMHD} second column). Additionally, one may quickly notice that the shocked plasma is not flat. Both phenomena are due to the magnetic force triggered by the strengthened toroidal magnetic component in the shocked plasma. As in both simulation setups the emission region is sufficiently magnetized, this magnetic force, which is in the radial direction, pinches the shocked plasma, squeezing the central jet ($r'<1$) while expanding outwards ($r'>1.5$) radially (Figure \ref{figure:RMHD}). Due to the frozen-in magnetic field lines, the squeezed central jet plasma has stronger poloidal component in a bullet shape. Meanwhile, the expansion traces a curved shocked plasma far downstream the shock front. The strength of this radial magnetic force is proportional to the local magnetization. From Figure \ref{figure:RMHD}, we can clearly see that the poloidal enhancement is stronger and the shape of the shocked plasma changes more in the high magnetization case than the low magnetization case.


Comparing the low and high magnetization cases, several differences are obvious. One is the magnetic field enhancement in the shocked plasma. Clearly, the toroidal component is more increased and the poloidal component is less enhanced in the low magnetization case. Given that our initial magnetic morphology has a higher poloidal component, we expect a stronger magnetic morphology change in the low magnetization case. The other difference is that the volume of the shocked plasma with enhanced thermal energy and higher velocity is much larger in the low magnetization case. Then, based on our particle injection method, one should expect that in the low magnetization environment both protons and electrons are distributed in a larger region where the magnetic field morphology is also different from the initial conditions. This should increase the polarization variability as well.

\subsection{MeV Polarized Variability}

\begin{figure*}[ht]
\parbox{\linewidth}{
\centering
\includegraphics[width=.48\textwidth]{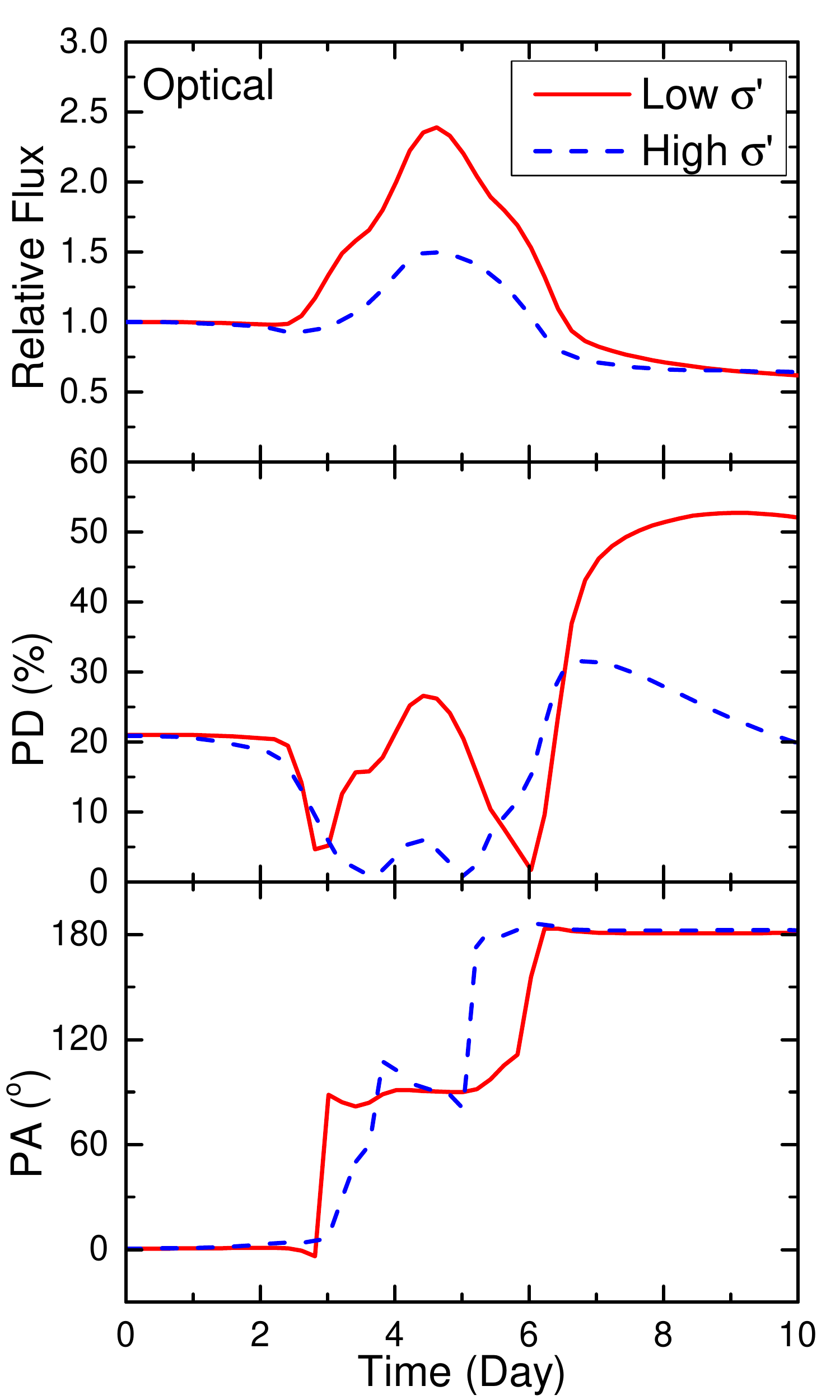}
\includegraphics[width=.48\textwidth]{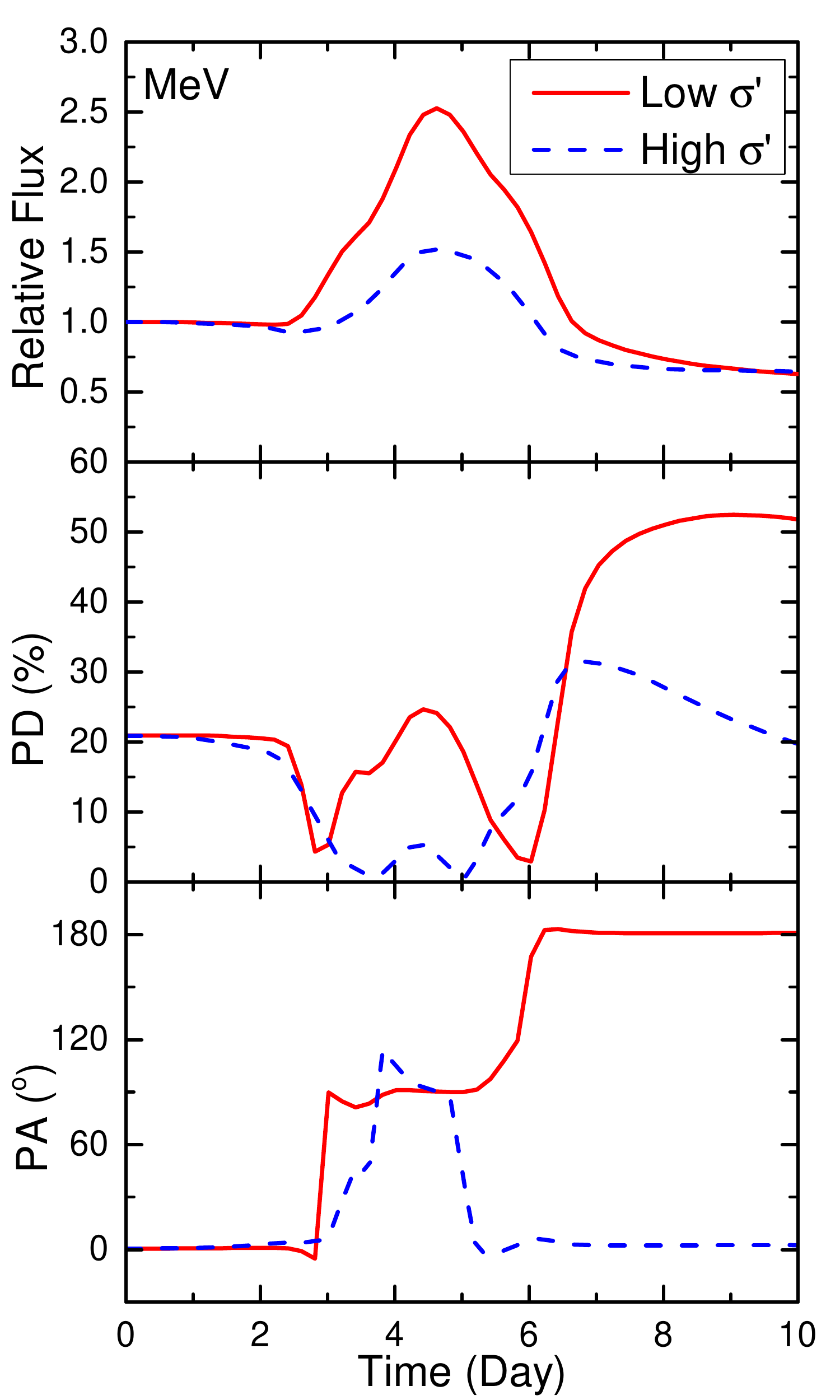}
}
\caption{Light curves, temporal polarization degree and angle curves in the optical and MeV $\gamma$-ray bands for the RMHD simulations in Figure \ref{figure:RMHD}. Left column is the optical polarized variability, and the right column is the MeV counterparts. The upper panels are the relative flux, with the initial state normalized to $1.0$ for both optical and MeV bands. The middle panels are the polarization degree in percentage. And the lower panels are the polarization angle in degrees. The red solid curves are for the low magnetization environment, and the blue dash curves are for the high magnetization environment. The total simulation time in the observer's frame is $\sim 10$ days, with the major flare duration $\sim 4.5$ days. All flux and polarization quantities are in the observer's frame. \label{figure:variability}}
\end{figure*}

The general polarized variability follows well with the magnetic evolution during shock propagation. Here we only plot the MeV light curves and polarization signatures. This is because the X-ray band emission should have a nontrivial contribution from the cascading pair synchrotron, which may alter the general polarization patterns. For higher energies, the general behaviors should be almost identical to the MeV band. Before the shock enters the emission region, since the initial magnetic morphology has a stronger poloidal component than the toroidal component, and the LOS is at $90^{\circ}$ from the jet axis in the comoving frame, the polarization angle stays at 0 along the jet axis with $20\%$ polarization degree. When the shock moves in, the toroidal component is strongly compressed. The same region also has more nonthermal particles due to the shock acceleration, leading to a flare in the light curve \ref{figure:variability}. Therefore, the toroidal contribution quickly takes over the poloidal contribution, and the polarization angle moves toward $90^{\circ}$. When the shock leaves the emission region, since the toroidal component is frozen in the shocked plasma, the downstream toroidal component is stretched and weakened due to the shock propagation. On the other hand, the poloidal component is hardly affected by the shock propagation. Therefore, the polarization reverts to an even more poloidal dominating situation than the initial state. Also due to the weakened toroidal magnetic field, the flux slightly decreases after the shock.

We can see that the optical and MeV light curves and polarization signatures are nearly identical. This is consistent with \cite{Zhang16b}, where they find that if the light crossing time scale is comparable to or longer than the proton cooling time scale, then the corresponding high-energy polarized variability should be similar to optical counterparts. The only difference between the optical and MeV curves in Figure \ref{figure:variability} is that in the high magnetization case the MeV polarization angle does not complete a polarization angle swing. This is because a complete $180^{\circ}$ polarization angle swing requires strong shock compression in the toroidal component, but in the high magnetization case the shock compression is not strong enough. In the optical band, since the shock injected nonthermal electrons strictly follow the magnetic compression at the shock front, the toroidal contribution is just enough to push a complete polarizaiton angle swing. However, the nonthermal protons are proportional to the thermal energy distribution, which covers a larger downstream region where the toroidal component is less dominating (Figure \ref{figure:RMHD}), the toroidal enhancement during the shock is not sufficient to push a complete polarization angle swing.

Comparing the low and high magnetization cases, we can easily see that the low magnetization case shows stronger flares and larger polarization variations in both optical and MeV bands. This is straightforward, as the shock compression is less efficient in a more magnetized environment. An interesting feature is that after the shock leaves the region, the polarization degree stays at a high level in the low magnetization case, but in the high magnetization case it can revert back to the initial value. This is because the magnetic force tries to restore a force-balance state. After the shock leaves the emission region, the poloidal component is strong in the central spine. This results in an outward radial magnetic force. If the emission region is sufficiently magnetized, this force can quickly expand the central spine and restore a force-free magnetic field topology, similar to the initial state. However, in the low magnetization environment, this force is relatively weak, thus the dominating poloidal component in the central spine cannot be relieved quickly, leading to a high polarization degree after the shock. Therefore, the restoration of the polarization signatures can help to diagnose the magnetization in the emission region.

\section{Implication on Observation}

Combining multi-wavelength polarimetry and neutrino detection, we can pinpoint the dominating radiation mechanisms in all high-energy bands. In a pure leptonic model such as Model 1, we do not expect any neutrino detection, and the high-energy polarization should be at most half as polarized as the optical bands. If one observes a decrease of the polarization towards higher energies, it infers the transition from SSC to EC processes. However, if one finds that the X-ray is similarly polarized as the optical band but the $\gamma$-ray polarization is much weaker, then it indicates a IC-dominated model with a cascading pair synchrotron component as in Model 2. We also expect to see neutrinos in such blazars. If one detects similar polarization degree in both X-ray and $\gamma$-ray bands at the level of the optical polarization, then the high-energy spectral component is dominated by the PS mechanism. However, PS models with different magnetic field strengths predict similar high-energy spectral polarization and neutrino flux. Therefore, we need time-dependent polarization signatures to distinguish them.

Two caveats, however, need to be mentioned. One is that if the blazar is an FSRQ that has a strong thermal component (the ``big blue bump''), then one should use the infrared to red polarization degree as the optical polarization to avoid thermal contamination. The second caveat is that if the blazar is a high-synchrotron-peaked so that the primary electron synchrotron extends to X-ray bands, then only the $\gamma$-ray polarization constraints can be employed.


Based on our MHD-integrated time-dependent polarization simulation, we can observe that the PS model with moderate magnetization factor ($\sigma'\sim 0.1$) predicts much stronger polarization variations in both degree and angle then the high magnetization case ($\sigma'\sim 1$). This applies to both optical and MeV $\gamma$-ray bands. We expect $\sim 20\%$ polarization degree during the flare and $> 40\%$ after the flare for the low magnetization environment. For the high magnetization case, the polarization degree stays below $\sim 30\%$ throughout the simulation, with $<10\%$ during the flare. Given that the TXS~0506+056 event has an observed optical polarization degree about $7\%$ during the flare \citep{IceCube18a}, it is likely that the emission zone has a high magnetization.

A very interesting phenomenon is that the MeV band can also present polarization angle swings similar to the optical band. Recent observations have found many optical polarization swings in blazars, and mostly they are accompanied by strong $\gamma$-ray flares \citep[e.g.,][]{Blinov16,Angelakis16,Blinov18}. In general, both random walks due to turbulence and large-scale magnetic field variations due to shocks or magnetic instabilities can explain these events \citep[e.g.,][]{Marscher14,Zhang16a,Tavecchio18}. However, in hadronic blazars, the proton cooling time scale is generally much longer than the random walk time scale, which is determined by the magnetic coherent length scale in the turbulence. As a result, the slow proton cooling should smooth out the stochastic behaviors of high-energy polarization signatures. Therefore, detection of high-energy polarization angle swing infers a major variation of the blazar zone magnetic morphology.

\section{Summary and Discussions}

In this paper, we predict the high-energy spectral polarization degree as well as their corresponding magnetic field strength and radiation mechanisms, based on detailed one-zone steady state spectral modeling. We find that the PS-dominated models generally predict higher polarization degree than the IC-dominated models across the entire high-energy spectral component. Also we find that the high-energy spectral polarization degree together with the neutrino detection can pinpoint the radiation mechanisms and magnetic field strength in the emission region. Additionally, we evaluate the temporal behaviors of multi-wavelength polarization signatures in PS-dominated models based on combined RMHD and polarized radiation transfer simulations. Our results suggest that the high-energy light curves and polarization signatures are very similar to the optical counterpart, except that the polarization signatures are slightly less variable. An interesting phenomenon that our work predicts is that the polarization signature in the high-energy band may show similar behaviour to the optical one, with polarization angle swings symptom of strong changes in the magnetic field morphology. However, this feature can only be observed with dedicated high-energy polarimeters with good temporal resolution. Furthermore, the time-dependent multi-wavelength polarimetry can shed light on the magnetic field evolution in the emission region. Future high-energy polarimeters such as {\it IXPE} and {\it AMEGO} are promising tools to test these features and put unprecedented constraints on the blazar emission, especially during their active phases.

Our semi-analytical study and numerical simulations aim at identifying key observational features and predictions of blazars with consistent a description of the underlying physics. However, in practice, there exist several important physical quantities that may impact our predictions. The first one is the size of the emission region. As shown in Table \ref{table:sed}, the total jet power in the PS-dominated model is considerably higher than the Eddington luminosity, 
\begin{equation}
L_{\rm Edd} = 4\pi GMm_pc/\sigma_T \sim 1.2\times 10^{47} M_9~erg\,s^{-1}~~.
\end{equation}
In fact, we can easily estimate the magnetic power in our hadronic model in section 3,
\begin{equation}
L_B=4\pi R^{\prime 2} c u'_B \Gamma_{obs}^2\sim 4\times 10^{49}~\rm{erg\,s^{-1}}~~,
\end{equation}
which by itself already exceeds the Eddington luminosity. Although occasionally it is possible that the emission region power can be higher than the Eddington luminosity, in general one expects that the former should be smaller than or comparable to the latter. Then the size of the emission region should be at least one order smaller than our numerical simulation parameters (Table \ref{table:RMHD}). In this case, however, the proton cooling time is much longer than the light crossing time. This means that for any jet energy dissipation and particle acceleration processes that take place in the emission region, the newly accelerated protons can linger over the entire duration of the flare. \cite{Zhang16b} has argued that these slow-cooling protons may result in nearly steady high-energy polarization signatures during flares; however, they assume that the magnetic field can revert to its initial state immediately after the passage of the dissipation process, which is not the case based on our RMHD simulations. On the other hand, for a large emission region such as in this paper, our results are consistent with \cite{Zhang16b}. Therefore, we suggest that a full description of the time-dependent high-energy polarization requires at least the combination of fluid dynamics, radiative cooling, and polarized radiation transfer altogether.

Magnetic instabilities and/or turbulence can also efficiently dissipate jet energy to accelerate particles through magnetic reconnection events \citep[e.g.,][]{Giannios09}. In particular, recent particle-in-cell (PIC) simulations have demonstrated that reconnection can accelerate protons as well \citep{Guo16}. Our spectral modeling parameters suggest that the magnetization factor in the PS-dominated models is generally between $0.1-1$. This indicates that hadronic blazar flares may be even preferably triggered by magnetic energy dissipation rather than shocks. Several previous works have studied the optical polarization signatures under kink instability and magnetic reconnection \citep{Zhang17,Nalewajko17,Zhang18}. However in these cases the magnetic field and nonthermal particles are co-evolving, which require careful treatment of the radiative cooling as well.

The last issue is that all our calculations assume a laminar setup. In practice, turbulence may widely exist in the blazar zone. Given that the gyroradius of the high-energy protons in the hadronic model can be much larger than that of the electrons, if the magnetic coherent length scale is smaller than the proton gyroradius, then protons can experience more turbulent field than electrons, leading to lower high-energy polarization degree than the optical bands. However, notice that on average the turbulent magnetic field should also partially cancel out the optical polarization. Therefore, we expect that overall the polarization degree should be similar in optical and high-energy bands for hadronic models, but optical bands may exhibit some fast high polarization spikes due to the small magnetic coherent lengths. Such stochastic behaviors of the optical polarization have been found in some leptonic models such as \cite{Marscher14,Tavecchio18}.

\acknowledgments{We acknowledge helpful discussions with B. Dingus. HZ acknowledges support from Fermi Guest Investigator program Cycle 10, grant number 80NSSC17K0753, and Cycle 11, grant number 80NSSC18K1723. KF acknowledges the support of a NASA Hubble Fellowship Program  Einstein fellowship.  HL acknowledges the support by the LANL/LDRD program, NASA/ATP program,  and the DoE/OFES program. HZ and DG acknowledge support from the NASA ATP NNX17AG21G and the NSF AST-1816136. MB acknowledges support by the South African Research Chairs Initiative (SARChI) of the Department of Science and Technology and the National Research Foundation\footnote{Any opinion, finding and conclusion or recommendation expressed in this material is that of the authors and the NRF does not accept any liability in this regard.} of South Africa.}

\clearpage


\begin{thebibliography}{}

\bibitem[Abdo et al.(2010)]{Abdo10} Abdo, A.~A., Ackermann, M., Ajello, M., et al.\ 2010, \nat, 463, 919 

\bibitem[Abeysekara et al.(2018)]{2018ApJ...861L..20A} Abeysekara, A.~U., Archer, A., Benbow, W., et al.\ 2018, \apjl, 861, L20 

\bibitem[Ackermann et al.(2016)]{Ackermann16} Ackermann, M., Anantua, R., Asano, K., et al.\ 2016, \apjl, 824, L20

\bibitem[Ansoldi et al.(2018)]{MAGIC18} Ansoldi, S., Antonelli, L.~A., Arcaro, C., et al.\ 2018, \apj, 863, L10.

\bibitem[Ajello et al.(2014)]{Ajello14} Ajello, M., Romani, R.~W., Gasparrini, D. et al.\ 2014 \apj 780, 73

\bibitem[Angelakis et al.(2016)]{Angelakis16} Angelakis, E., Hovatta, T., Blinov, D., et al.\ 2016, \mnras, 463, 3365 

\bibitem[Atoyan \& Dermer(2001)]{2001PhRvL..87v1102A} Atoyan, A., \& Dermer, C.~D.\ 2001, Physical Review Letters, 87, 221102 

\bibitem[Bednarek \& Protheroe(1999)]{1999MNRAS.302..373B} Bednarek, W., \& Protheroe, R.~J.\ 1999, \mnras, 302, 373 

\bibitem[Blinov et al.(2016)]{Blinov16} Blinov, D., Pavlidou, V., Papadakis, I.~E., et al.\ 2016, \mnras, 457, 2252 

\bibitem[Blinov et al.(2018)]{Blinov18} Blinov, D., Pavlidou, V., Papadakis, I., et al.\ 2018, \mnras, 474, 1296 

\bibitem[Boettcher et al.(2012)]{Boettcher12} Boettcher, M., Harris, D.~E., \& Krawczynski, H.\ 2012, Relativistic Jets from Active Galactic Nuclei, by M.~Boettcher, D.E.~Harris, ahd H.~Krawczynski, 425 pages.~ Berlin: Wiley, 2012,  

\bibitem[B{\"o}ttcher et al.(2013)]{Boettcher13} B{\"o}ttcher, M., Reimer, A., Sweeney, K., \& Prakash, A.\ 2013, \apj, 768, 54 

\bibitem[Cerruti et al.(2015)]{Cerruti15} Cerruti, M., Zech, A., Boisson, C., \& Inoue, S.\ 2015, \mnras, 448, 910 

\bibitem[Cerruti et al.(2019)]{Cerruti19} Cerruti, M., Zech, A., Boisson, C., et al.\ 2019, \mnras, 483, L12

\bibitem[Chen et al.(2014)]{Chen14} Chen, X., Chatterjee, R., Zhang, H., et al.\ 2014, \mnras, 441, 2188 

\bibitem[Dermer et al.(1992)]{Dermer92} Dermer, C. D., et al., 1992, A\&A, 256, L27

\bibitem[Dermer \& Menon(2009)]{DermerBook} Dermer, C.~D., \& Menon, G.\ 2009, High Energy Radiation from Black Holes: Gamma Rays, Cosmic Rays, and Neutrinos by Charles D.~Dermer and Govind Menon.~Princeton Univerisity Press, November 2009.,  
\bibitem[Dermer et al.(2012)]{2012ApJ...755..147D} Dermer, C.~D., Murase, K., \& Takami, H.\ 2012, \apj, 755, 147 


\bibitem[Dermer et al.(2014)]{Dermer14} Dermer, C.~D., Murase, K., \& Inoue, Y.\ 2014, Journal of High Energy Astrophysics, 3, 29 

\bibitem[Diltz et al.(2015)]{Diltz15} Diltz, C., B{\"o}ttcher, M., \& Fossati, G.\ 2015, \apj, 802, 133 


\bibitem[Eichmann et al.(2012)]{2012ApJ...749..155E} Eichmann, B., Schlickeiser, R., \& Rhode, W.\ 2012, \apj, 749, 155 

\bibitem[Ghisellini \& Tavecchio(2008)]{GT08} Ghisellini, G. \& Tavecchio, F.\ 2008, \mnras, 387, 1669

\bibitem[Gao et al.(2019)]{Gao19} Gao, S., Fedynitch, A., Winter, W., et al.\ 2019, Nature Astronomy, 3, 88.

\bibitem[Giannios et al.(2009)]{Giannios09} Giannios, D., Uzdensky, D.~A., \& Begelman, M.~C.\ 2009, \mnras, 395, L29 

\bibitem[Guan et al.(2014)]{Guan14} Guan, X., Li, H., \& Li, S.\ 2014, \apj, 781, 48 

\bibitem[Guo et al.(2016)]{Guo16} Guo, F., Li, X., Li, H., et al.\ 2016, \apjl, 818, L9 

\bibitem[IceCube Collaboration(2013)]{IceCube13} IceCube Collaboration 2013, Science, 342, 1242856 

\bibitem[IceCube Collaboration (2018a)]{IceCube18a} IceCube Collaboration, 2018a, Science 361, 6398

\bibitem[IceCube Collaboration (2018b)]{IceCube18b} IceCube Collaboration, 2018b, Science, doi:10.1126/science.aat2890

\bibitem[Kelner \& Aharonian(2008)]{Kelner08} Kelner, S.~R., \& Aharonian, F.~A.\ 2008, \prd, 78, 034013 

\bibitem[Keivani et al.(2018)]{Keivani18} Keivani, A., Murase, K., Petropoulou, M., et al.\ 2018, \apj, 864, 84.

\bibitem[Laing(1980)]{Laing80} Laing, R.~A.\ 1980, \mnras, 193, 439 

\bibitem[Li \& Kusunose(2000)]{Li00} Li, H., \& Kusunose, M.\ 2000, \apj, 536, 729 

\bibitem[Li \& Li(2003)]{Li03} Li, S., \& Li,H., 2003, Los Alamos National Lab. Tech. Rep. LA-UR-03-8935

\bibitem[Liu et al.(2014)]{Liu14} Liu, R.-Y., Wang, X.-Y., Inoue, S., Crocker, R., \& Aharonian, F.\ 2014, \prd, 89, 083004 

\bibitem[Liu et al.(2018)]{Liu18} Liu, R.-Y., Wang, K., Xue, R., et al.\ 2018, arXiv:1807.05113 

\bibitem[Mannheim \& Biermann(1992)]{Mannheim92a} Mannheim, K., \& Biermann, P.~L.\ 1992, \aap, 253, L21 

\bibitem[Mannheim et al.(1992)]{Mannheim92b} Mannheim, K., Stanev, T., \& Biermann, P.~L.\ 1992, \aap, 260, L1 

\bibitem[Maraschi et al.(1992)]{Maraschi92} Maraschi, L., et al., 1992, ApJ, 397, L5

\bibitem[Marscher \& Gear(1985)]{Marscher85} Marscher, A. P. \& Gear, W. K., 1985, ApJ, 298, 114

\bibitem[Marscher(2014)]{Marscher14} Marscher, A.~P.\ 2014, \apj, 780, 87 

\bibitem[McEnery(2017)]{McEnery17} McEnery, J.~E.\ 2017, AAS/High Energy Astrophysics Division \#16, 16, 103.13 

\bibitem[Mizuno et al.(2009)]{Mizuno09} Mizuno, Y., Lyubarsky, Y., Nishikawa, K.-I., \& Hardee, P.~E.\ 2009, \apj, 700, 684 

\bibitem[M{\"u}cke \& Protheroe(2001)]{Mucke01} M{\"u}cke, A., \& Protheroe, R.~J.\ 2001, Astroparticle Physics, 15, 121 

\bibitem[M{\"u}cke et al.(2003)]{2003APh....18..593M} M{\"u}cke, A., Protheroe, R.~J., Engel, R., Rachen, J.~P., \& Stanev, T.\ 2003, Astroparticle Physics, 18, 593 


\bibitem[Murase \& Nagataki(2006)]{2006PhRvD..73f3002M} Murase, K., \& Nagataki, S.\ 2006, \prd, 73, 063002 

\bibitem[Murase et al.(2013)]{Murase13} Murase, K., Ahlers, M., \& Lacki, B.~C.\ 2013, \prd, 88, 121301 

\bibitem[Murase et al.(2018)]{Murase18} Murase, K., Oikonomou, F., \& Petropoulou, M.\ 2018, \apj, 865, 124.

\bibitem[Nalewajko(2017)]{Nalewajko17} Nalewajko, K.\ 2017, Galaxies, 5, 64 

\bibitem[Padovani et al.(2018)]{Padovani18} Padovani, P., Giommi, P., Resconi, E., et al.\ 2018, \mnras, 480, 192.

\bibitem[Padovani et al.(2019)]{Padovani19} Padovani, P., Oikonomou, F., Petropoulou, M., Giommi, P., \& Resconi, E.\ 2019, \mnras, 484, L104 

\bibitem[Paiano et al.(2018)]{Paiano18} Paiano, S., Falomo, R., Treves, A., \& Scarpa, R.\ 2018, \apjl, 854, L32 

\bibitem[Paliya et al.(2018)]{Paliya18} Paliya, V.~S., Zhang, H., B{\"o}ttcher, M., et al.\ 2018, \apj, 863, 98.

\bibitem[Petropoulou et al.(2015)]{Petropoulou15} Petropoulou, M., Dimitrakoudis, S., Padovani, P., Mastichiadis, A., \& Resconi, E.\ 2015, \mnras, 448, 2412 

\bibitem[Reimer et al.(2018)]{Reimer18} Reimer, A., Boettcher, M., \& Buson, S.\ 2018, arXiv:1812.05654 

\bibitem[Rybicki \& Lightman(1979)]{Rybicki79} Rybicki, G.~B., \& Lightman, A.~P.\ 1979, New York, Wiley-Interscience, 1979.~393 p.,  

\bibitem[Scarpa \& Falomo(1997)]{Scarpa97} Scarpa, R., \& Falomo, R.\ 1997, \aap, 325, 109 

\bibitem[Sikora et al.(1994)]{Sikora94} Sikora, M., et al., 1994, \apj, 421, 153

\bibitem[Tamborra et al.(2014)]{Tamborra14} Tamborra, I., Ando, S., \& Murase, K.\ 2014, \jcap, 9, 043 

\bibitem[Tavecchio \& Ghisellini(2015)]{Tavecchio15} Tavecchio, F., \& Ghisellini, G.\ 2015, \mnras, 451, 1502 

\bibitem[Tavecchio et al.(2018)]{Tavecchio18} Tavecchio, F., Landoni, M., Sironi, L., \& Coppi, P.\ 2018, \mnras, 480, 2872 

\bibitem[Weisskopf et al.(2016)]{Weisskopf16} Weisskopf, M.~C., Ramsey, B., O'Dell, S., et al.\ 2016, \procspie, 9905, 990517 

\bibitem[Werner et al.(2018)]{Werner18} Werner, G.~R., Uzdensky, D.~A., Begelman, M.~C., Cerutti, B., \& Nalewajko, K.\ 2018, \mnras, 473, 4840 

\bibitem[Zhang \& B{\"o}ttcher(2013)]{Zhang13} Zhang, H., \& B{\"o}ttcher, M.\ 2013, \apj, 774, 18 

\bibitem[Zhang et al.(2016a)]{Zhang16a} Zhang, H., Deng, W., Li, H., \& B\"ottcher, M., 2016a, ApJ, 817, 63

\bibitem[Zhang et al.(2016b)]{Zhang16b} Zhang, H., Diltz, C., \& B{\"o}ttcher, M.\ 2016b, \apj, 829, 69 

\bibitem[Zhang et al.(2017)]{Zhang17} Zhang, H., Li, H., Guo, F., \& Taylor, G.\ 2017, \apj, 835, 125 

\bibitem[Zhang et al.(2018)]{Zhang18} Zhang, H., Li, X., Guo, F., et al.\ 2018, \apj, 862, L25.

\end{thebibliography}
\end{document}